\documentclass[conference]{IEEEtran}
\IEEEoverridecommandlockouts

\usepackage{pgfplots}

\usepackage{cite}
\usepackage{amsmath,amssymb,amsfonts}
\usepackage{algorithmic}
\usepackage{graphicx}
\usepackage{algorithm,algorithmic}
\usepackage{hyperref}
\hypersetup{hidelinks=true}
\usepackage{textcomp}

\usepackage{comment}
\usepackage{booktabs}
\usepackage[table]{xcolor} 
\usepackage{colortbl} 

\definecolor{low}{RGB}{198, 239, 206}    
\definecolor{medium}{RGB}{255, 235, 156} 
\definecolor{high}{RGB}{255, 199, 206}   

\usepackage{cite}
\usepackage{amsmath,amssymb,amsfonts}
\usepackage{algorithmic}
\usepackage{graphicx}
\usepackage{textcomp}
\usepackage{xcolor}
\usepackage{subfigure}
\usepackage{caption, subcaption}
\usepackage{amsmath,graphicx}
\usepackage{textcomp}
\usepackage{amsmath,amssymb,amsfonts}
\usepackage{mathtools}
\usepackage{amsmath,graphicx, multirow, hyperref}
\usepackage{subcaption, subfigure}
\usepackage{graphicx,subfigure}

\newcommand{\cognospeak}{CognoMemory}

\newcommand{\dementiabank}{DementiaBank}
\newcommand{\bart}{BART}

\newcommand{\roberta}{RoBERTa}
\newcommand{\distilbert}{DistilBERT}

\newcommand{\egemaps}{eGeMAPS}
\newcommand{\compare}{ComParE}

\newcommand{\Whisper}{Whisper}
\newcommand{\WV}{Wav2Vec 2.0}
\newcommand{\Nemo}{NeMo}

\newcommand{\SentenceBERT}{Sentence-BERT}
\newcommand{\WordVec}{Word2Vec}
\newcommand{\FastText}{FastText}

\usepackage{pgfplotstable}
\usepackage[table]{xcolor}
\usepackage{booktabs}
\usepackage{colortbl}
\usepackage{multirow}
\usepackage{caption}
\usepackage{tikz}
\pgfplotsset{compat=1.18}

\newcommand{\minWER}{0.3}
\newcommand{\maxWER}{0.65}

\newcommand{\heatmapcell}[1]{%
    \pgfmathsetmacro{\value}{#1}
    \pgfmathsetmacro{\percent}{100*(\value - \minWER)/(\maxWER - \minWER)}
    \edef\cellcolorval{\noexpand\cellcolor{red!\percent!white}}%
    \cellcolorval #1%
}

\newcommand{\heatmapcellPVALs}[1]{%
    \pgfmathsetmacro{\value}{#1}
    \pgfmathsetmacro{\percent}{100*\value}
    \ifdim \percent pt < 5 pt
        \pgfmathsetmacro{\setCellColor}{"green!30!white"}
    \else
        \pgfmathsetmacro{\setCellColor}{"red!50!white"}
    \fi
    \edef\cellcolorval{\noexpand\cellcolor{\setCellColor}}%
    \cellcolorval #1%
}

\newcommand{\sig}[1]{%
  #1%
  \pgfmathsetmacro{\pvalue}{#1}%
  \ifdim \pvalue pt < 0.001 pt\relax$^{***}$%
  \else\ifdim \pvalue pt < 0.01 pt\relax$^{**}$%
  \else\ifdim \pvalue pt < 0.05 pt\relax$^{*}$%
  \fi\fi\fi%
}

\def\BibTeX{{\rm B\kern-.05em{\sc i\kern-.025em b}\kern-.08em
    T\kern-.1667em\lower.7ex\hbox{E}\kern-.125emX}}
\begin{document}

\title{CCMAN: Cognitive Instability-Aware Cross-Modal Attention Network for Interpretable Temporal Biomarkers of Verbal Fluency Speech\\
}

\author{\IEEEauthorblockN{Madhurananda Pahar$^1$,
Caitlin Illingworth$^2$,
Dorota Braun$^2$,
Daniel Blackburn$^2$,
Heidi Christensen$^1$
}
\vspace{5pt}
\IEEEauthorblockA{$^1$School of Computer Science, University of Sheffield, Sheffield, S1 4DP, UK\\
\vspace{5pt}
$^2$Sheffield Institute for Translational Neuroscience (SITraN), University of Sheffield, Sheffield, S10 2HQ, UK\\
Email: \{m.pahar, c.illingworth, d.a.braun, d.blackburn, heidi.christensen\}@sheffield.ac.uk
}
}


\maketitle

\begin{abstract}

    Early detection of cognitive decline from speech offers a scalable and non-invasive alternative to conventional clinical assessment. Verbal fluency tasks are particularly informative, but most automated approaches aggregate features across an entire recording, overlooking temporal speech dynamics. We propose the Cognitive Instability-Aware Cross-Modal Attention Network (CCMAN), a transfer learning framework that learns task-agnostic cognitive speech representations from multiple memory-probing tasks before fine-tuning on a minute-long semantic and phonemic verbal fluency task. CCMAN integrates semantic, acoustic, and linguistic information through bidirectional cross-attention, gated multimodal fusion, and transformer-based temporal modelling to derive interpretable biomarkers of cognitive decline. Experiments were conducted on 165.44 hours of speech from 843 participants (498 healthy controls, 245 with mild cognitive impairment, and 100 with dementia). CCMAN achieved Macro-$F_1$ scores of 0.81 and 0.59 for binary and multiclass semantic fluency classification, and 0.77 and 0.53 for phonemic fluency, consistently outperforming strong static and temporal baselines. 
    Statistical analyses showed that semantic drift variance and pause variance, but not mean semantic drift, were significantly elevated in both MCI and dementia relative to healthy controls, while pause duration increased progressively over the task with the steepest slope in dementia, supporting global and progressive temporal speech instability as interpretable biomarkers.
    Evaluation on the independent PROCESS-2 benchmark further demonstrated the generalisability of the proposed framework, improving the baseline Macro-$F_1$ by up to 9\%. 
    These findings support temporal speech instability as a dynamic speech biomarker for robust, interpretable, and generalisable early detection of cognitive decline.

\end{abstract}


\section{Introduction}

Healthy older adults often experience mild age-related decreases in their memory and general cognition, though the impact on functioning is negligible \cite{murman2015impact}. When cognitive ability declines to significantly below age group norms, these changes may indicate mild cognitive impairment (MCI) \cite{davis2018estimating}, characterised by deficits in one or more cognitive domains such as memory, learning, reasoning, attention, and language \cite{rosenberg2013association, petersen2004mild}.
Approximately half of individuals living with MCI progress to dementia \cite{prestia2013prediction}, an umbrella term for neurodegenerative disorders, most commonly Alzheimer’s disease (AD). Dementia is the leading cause of death in the UK, and a major cause of dependence in older populations globally, resulting in memory loss, executive dysfunction, and language impairment \cite{davis2018estimating, rosenberg2013association}. Early detection is essential to enable timely intervention, optimise care planning, and facilitate clinical trial recruitment \cite{robin2021using}, underscoring the need for smart remote technologies that support accurate and timely diagnosis \cite{gauthier2021world}.

Instabilities or variances in speech and linguistic biomarkers, such as pauses, word count, repetitions or category switches, are traditional scoring methods that provide valuable but coarse indicators of cognitive performance \cite{he2025automated, segkouli2025linguistic}. 
Traditional neuropsychological assessments such as the semantic and phonemic verbal fluency tasks have long been used to identify early signs of cognitive decline by measuring the ability to retrieve and produce words under time constraint \cite{nutter2008verbal, macoir2022use, frankenberg2021verbal}. 
However, speech production is a dynamic process involving continuous planning, lexical retrieval, and articulation, and static aggregates often fail to capture this temporal biomarker \cite{vigo2022speech, li2025associations}.

\begin{figure*}[h]
\centering
\includegraphics[width=\linewidth]{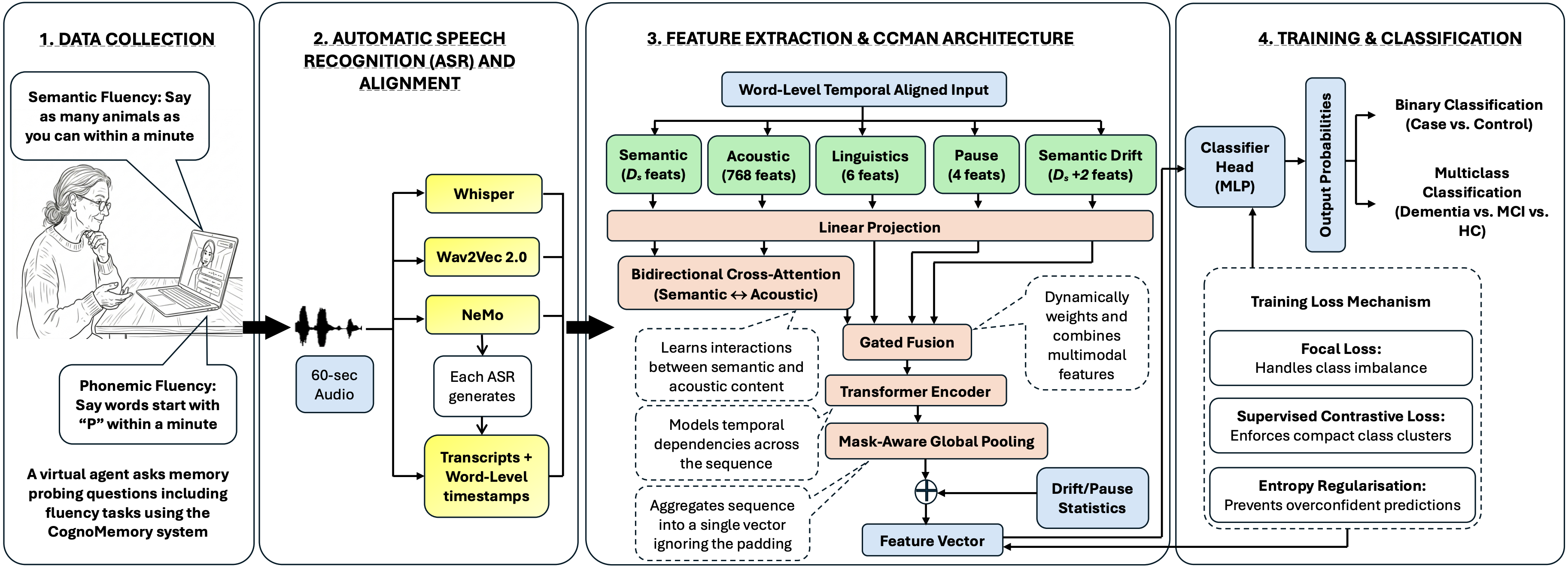}
\caption{
Overview of the proposed Cognitive Instability-Aware Cross-Modal Attention Network (CCMAN) framework.
(1) Participants complete semantic and phonemic fluency tasks via the \cognospeak{} virtual agent interface. 
(2) Speech is transcribed using multiple ASR systems (Whisper \cite{radford2022whisper}, \WV{}\cite{baevski2020wav2vec}, and NVIDIA \Nemo{} \cite{kuchaiev2019nemo}), and word-level timestamp alignment is performed. 
(3) Multimodal features (semantic embeddings, acoustic representations, linguistic descriptors, pause duration, and semantic drift measures) are extracted from each word (Section \ref{subsec:featExrct}), and are projected into a shared latent space and integrated via bidirectional cross-attention and gated fusion, followed by a Transformer encoder and mask-aware global pooling to obtain an assessment-level representation (Section \ref{subsec:CCMANarch}). 
(4) The resulting feature vector, augmented with global drift and pause statistics, is used to optimise CCMAN using focal loss, supervised contrastive loss, and entropy regularisation, and finally used for binary or multiclass classification (Section \ref{subsec:trainingProc}).
}
\label{fig:overall_arch}
\end{figure*}

\cognospeak{} is an innovative online AI tool which identifies early cognitive decline, designed to reduce pressure on healthcare systems worldwide, such as the NHS in the UK \cite{pahar2025cognospeak, pahar2025mutlimodalfusion, yaointerspeech26, young2025can, pahar2025cognospeakWiley, illingworth2025developing}, by extracting static (aggregated) features from the audio recordings collected from participants answering multiple memory-probing questions, including fluency tasks. 
Although it exhibits strong performance, no interpretable speech characteristics were generated from our prior research. 
In this work, however, we investigate not only the fluency tasks to improve classification performance but, more importantly, identify and interpret temporal speech-based biomarkers for cognitive decline. 
To meet these research objectives, we introduce CCMAN, an interpretable \textit{Cognitive Instability-Aware Cross-Modal Attention Network}, a unified architecture that models word-level sequential speech representations using both semantic and phonemic fluency tasks for early detection of cognitive decline. We further demonstrate a hierarchical ablation framework that systematically quantifies the contributions of semantic drift, cross-modal attention, and entropy regularisation, demonstrating consistent performance gains over strong static and temporal baselines in both binary and multi-class classification settings. 
Most importantly, we provide a comparative analysis between global aggregated features and temporal dynamics, showing that intra-recording instability measures (e.g., drift variance, pause progression, attention entropy) offer statistically significant and clinically interpretable discriminative biomarkers across dementia, MCI, and healthy control (HC) groups.

\section{Previous Work}

Automated speech analysis provides a scalable, non-invasive approach to detecting cognitive decline. Acoustic and linguistic features from spontaneous or constrained speech, such as pause frequency, speech rate, lexical diversity, syntactic complexity, and semantic coherence, have shown promising discrimination between MCI, AD, and HC \cite{cohen2025speech, vigo2022speech, ter2023accuracy, vincze2022linguistic, garcia2024unveiling}. More recently, deep learning and foundation-model approaches using ASR-derived and multimodal embeddings have achieved competitive performance in dementia classification benchmarks \cite{agbavor2024multilingual, braun2024infusing, khatri2024alzheimer, chlasta2025enhancing}.
Temporal speech markers, including silent pause duration, hesitation frequency, and word transition timing, reflect lexical retrieval and executive processes and are associated with MCI and AD \cite{imre2022temporal, vincze2022linguistic}. However, such analyses are often observational, cohort-limited, or based on manually derived statistics, and are rarely incorporated into end-to-end predictive frameworks.

Despite progress in multimodal modelling, many methods rely on static or aggregated features, limiting sensitivity to intra-utterance dynamics and interpretability \cite{yang2022deep, ugwu2025temporal}. Consequently, the temporal dynamics of speech, such as pause trajectories, semantic drift, and lexical entropy progression, remain under-explored, and the interpretability of group-level temporal differences is limited \cite{zhu2022towards}, which we address in our study.

\section{Data}
\label{sec:data}

\begin{table*}[t]
\centering
\small
\caption{Dataset composition for transfer learning. Demographic characteristics, speech duration, and transcript statistics for the pre-training corpus (12 memory-probing tasks) and downstream fine-tuning corpus (semantic and phonemic verbal fluency tasks). Audio duration is reported in hours and transcript counts in thousands of words. Gender is reported as Male/Female/Undisclosed.}
\label{tab:dataset_summary_asr_group}
\setlength{\tabcolsep}{9pt}
\renewcommand\arraystretch{1.3}
\begin{tabular}{llcccccccc}
\toprule
\multirow{2}{*}{\textbf{Group}} & 
\multirow{2}{*}{\textbf{Diagnosis}} & 
\multirow{2}{*}{\textbf{\textit{N}}} & 
\textbf{Gender} & 
\textbf{Age} & 
\multirow{2}{*}{\textbf{Task}} & 
\textbf{Audio} & 
\textbf{Whisper} & 
\textbf{W2V2} & 
\textbf{NeMo} \\
 & & & \textbf{(M/F/U)} & \textbf{(Mean $\pm$ SD)} &  & \textbf{(hr)} & \textbf{($\times 10^3$)} & \textbf{($\times 10^3$)} & \textbf{($\times 10^3$)} \\
\midrule
\multirow{7}{*}{Case} & \multirow{2}{*}{Dementia} & \multirow{2}{*}{100} & \multirow{2}{*}{54 / 44 / 2} & \multirow{2}{*}{74.77 $\pm$ 7.66} & Pre-train & 15.23 & 89.24 & 93.89 & 85.83 \\
 & & & & & Fine-tune & 3.34 & 5.16 & 5.09 & 4.09 \\
\cline{2-10}
 & \multirow{2}{*}{MCI} & \multirow{2}{*}{245} & \multirow{2}{*}{138 / 102 / 5} & \multirow{2}{*}{73.36 $\pm$ 6.86} & Pre-train & 43.39 & 293.00 & 305.50 & 282.34 \\
 & & & & & Fine-tune & 8.18 & 15.17 & 14.84 & 12.66 \\
\cline{2-10}
 & \multirow{3}{*}{\textit{Subtotal}} & \multirow{3}{*}{\textit{345}} & \multirow{3}{*}{\textit{192 / 146 / 7}} & \multirow{3}{*}{\textit{73.77 $\pm$ 7.12}} & \textit{Pre-train} & \textit{58.62} & \textit{382.24} & \textit{399.39} & \textit{368.17} \\
 & & & & & \textit{Fine-tune} & \textit{11.52} & \textit{20.33} & \textit{19.94} & \textit{16.75} \\
\cline{6-10}
 & & & & & \textit{Total} & \textit{70.14} & \textit{402.57} & \textit{419.33} & \textit{384.92} \\
\midrule
\multirow{2}{*}{Control} & \multirow{2}{*}{HC} & \multirow{2}{*}{498} & \multirow{2}{*}{225 / 265 / 8} & \multirow{2}{*}{73.87 $\pm$ 4.97} & Pre-train & 78.67 & 577.32 & 605.11 & 566.62 \\
 & & & & & Fine-tune & 16.63 & 32.89 & 33.46 & 28.95 \\
\midrule
\multirow{3}{*}{\textbf{Total}} & \multirow{3}{*}{\textbf{All}} & \multirow{3}{*}{\textbf{843}} & \multirow{3}{*}{\textbf{417 / 411 / 15}} & \multirow{3}{*}{\textbf{73.83 $\pm$ 5.94}} & \textbf{Pre-train} & \textbf{137.30} & \textbf{959.56} & \textbf{1004.49} & \textbf{934.79} \\
 & & & & & \textbf{Fine-tune} & \textbf{28.14} & \textbf{53.22} & \textbf{53.39} & \textbf{45.70} \\
\cline{6-10}
 & & & & & \textbf{Total} & \textbf{165.44} & \textbf{1012.78} & \textbf{1057.88} & \textbf{980.49} \\
\bottomrule
\end{tabular}
\end{table*}

Data were collected using the \cognospeak{} system \cite{pahar2025cognospeak, pahar2025mutlimodalfusion}, in which a virtual conversational agent administered 14 memory-probing questions, including one semantic verbal fluency task (naming animals) and one phonemic verbal fluency task (producing words beginning with the letter \textit{P}), each lasting 60 seconds \cite{vaughan2018semantic} (Figure~\ref{fig:overall_arch}).
The dataset comprised 843 monolingual English-speaking participants aged over 50 years \cite{pahar2026can, illingworth2026assessing, pahar2026false}, including 100 individuals diagnosed with dementia, 245 with MCI, and 498 HC. The overall mean age was 73.83 $\pm$ 5.94 years, with no statistically significant differences in age across diagnostic groups.
To enable transfer learning, the dataset was divided according to task type. The remaining 12 memory-probing questions (excluding semantic and phonemic verbal fluency) were used for pre-training the proposed CCMAN encoder, while the two verbal fluency tasks served as the downstream fine-tuning task for cognitive status classification.

For each recording, transcripts were generated independently using the Whisper, Wav2Vec~2.0, and NeMo ASR systems. On manually annotated speech collected using the same \cognospeak{} platform, these ASR systems achieved an average word error rate (WER) of approximately 20\% \cite{pahar2026can, pahar2026false}. Although this WER is relatively high, it reflects the challenges of spontaneous, real-world clinical conversations involving older adults, including disfluencies, hesitations, variable recording conditions, and speech impairments associated with cognitive decline. Using multiple ASR systems further exposes the model to transcription variability, encouraging the learning of representations that are robust to ASR errors before being specialised for verbal fluency-based dementia classification.
As summarised in Table~\ref{tab:dataset_summary_asr_group}, the pre-training corpus consisted of 137.30 hours of spontaneous speech, corresponding to approximately 83\% of the available audio and this corpus yielded 959.56k, 1004.49k, and 934.79k words using Whisper, Wav2Vec 2.0, and NeMo, respectively. 
In contrast, the downstream fine-tuning corpus comprised 28.14 hours of semantic and phonemic verbal fluency recordings, generating 53.22k, 53.39k, and 45.70k words, respectively.
Overall, the complete dataset contains 165.44 hours of speech and over 1 million automatically transcribed words, making it one of the largest speech corpora collected for cognitive assessment.
Within the pre-training corpus, healthy controls contributed 78.67 hours (57.3\%) of speech, followed by participants with MCI (43.39 hours) and dementia (15.23 hours). The corresponding fine-tuning corpus contained 16.63 hours of healthy control recordings, 8.18 hours from participants with MCI, and 3.34 hours from participants with dementia. As expected, speech duration and transcript volume closely reflected the number of participants within each diagnostic group, with healthy controls contributing the largest proportion of recordings.

\section{Experimental Setup}
\label{sec:expSETUP}

\subsection{Multimodal Temporal Feature Representation}
\label{subsec:featExrct}

Each 
60-second verbal fluency 
recording is represented as a temporally aligned sequence of $T$ word-level tokens using ASR timestamps (Figure~\ref{fig:overall_arch}). 
For each token $t$, we construct:
\[
\mathbf{x}_t =
\left[
\mathbf{s}_t,
\mathbf{a}_t,
\mathbf{l}_t,
\mathbf{p}_t,
\mathbf{d}_t
\right]
\]

Where, $\mathbf{s}_t \in \mathbb{R}^{D_s}$ (semantic embeddings) are pretrained word representations (from \SentenceBERT{}, $D_s=384$ \cite{reimers2019sentence}; \WordVec{} or \FastText{}, $D_s=300$ \cite{mikolov2013efficient, bojanowski2017enriching}) encoding semantic similarity in a continuous space.
Acoustic embeddings ($\mathbf{a}_t \in \mathbb{R}^{768}$) are extracted from a pretrained \WV{} model; frame-level features are aligned to word boundaries and averaged, yielding contextualised phonetic–prosodic representations per token.
Linguistic descriptors ($\mathbf{l}_t \in \mathbb{R}^{6}$) from spaCy \cite{honnibal2020spacy} include word length, stopword indicator, content-word indicator (noun/verb/adjective/adverb), type–token ratio, content-word ratio, and stopword ratio, capturing lexical complexity and content–function balance.
Pause features ($\mathbf{p}_t \in \mathbb{R}^{4}$) encode inter-word temporal dynamics: raw pause duration, $\log(1+\text{pause})$, long-pause indicator ($>0.5$s), and first-order pause differences.
Semantic drift ($\mathbf{d}_t \in \mathbb{R}^{D_s+2}$) models local semantic transitions using consecutive embedding difference vectors, their L2 norm (drift magnitude), and cosine similarity (directional consistency).

\begin{table*}[h]
\setlength{\tabcolsep}{15pt}
\centering
\caption{CCMAN Fine-tuning configuration: frozen vs. trainable components.}
\label{tab:finetune_config}
\begin{tabular}{@{}lccc@{}}
\toprule
\textbf{Component} & \textbf{Pre-trained?} & \textbf{Fine-tuned?} & \textbf{Re-initialised?} \\
\midrule
Modality projections ($W_s, W_a, W_o$) & Yes & Frozen & No \\
Bidirectional cross-attention & Yes & Frozen & No \\
Transformer encoder & Yes & Frozen & No \\
Projection head ($\mathrm{MLP}_p$) & Yes & Trainable & No \\
Classifier hidden layers ($\mathrm{MLP}_c$ first layers) & Yes & Trainable & No \\
Classifier final layer & No & Trainable & Yes (per $C$) \\
\bottomrule
\end{tabular}
\end{table*}

\subsection{CCMAN Architecture}
\label{subsec:CCMANarch}
Our proposed CCMAN, Cognitive Instability-Aware Cross-Modal Attention Network, jointly models semantic, acoustic, and auxiliary conversational dynamics.

Given an input sequence $X \in \mathbb{R}^{B \times T \times D}$, features are partitioned into semantic ($D_s=384$ or $300$), acoustic ($D_a=768$), and auxiliary components $D_o=D_l+D_p+D_d$, where $D_l=6$ (linguistic cues), $D_p=4$ (pause features), and $D_d=D_s+2$ (drift features). Each modality is linearly projected into a shared latent space ($d=256$).

Bidirectional cross-modal attention captures interactions between semantic and acoustic streams:
\[
\tilde{S}=\mathrm{MHA}(S,A,A), \qquad
\tilde{A}=\mathrm{MHA}(A,S,S),
\]
with padding-aware masking. Attention entropy is computed to quantify cross-modal alignment dispersion.

Fusion is performed using a learned sigmoid gate:
\[
G=\sigma(\tilde{S}+\tilde{A}), \qquad
F=G\odot\tilde{S}+(1-G)\odot\tilde{A}+O,
\]
where $O$ denotes the projected auxiliary features. The attended semantic and acoustic representations are adaptively fused, augmented with the auxiliary features, and refined by a Transformer encoder to capture longer-range temporal dependencies.

Masked mean pooling produces a sequence-level representation $\mathbf{z}$. Global instability statistics (drift mean, drift variance, and pause variance) are computed over valid tokens and concatenated with $\mathbf{z}$ before a projection head generates a normalised 64-dimensional embedding. The resulting representation is passed to a classification head for diagnosis prediction.


\subsection{Transfer Learning Setup}

To further enhance generalisation and leverage additional data beyond the 60-second fluency tasks, we adopt a two-stage transfer learning strategy: pre-training on a larger auxiliary dataset followed by fine-tuning on the target fluency tasks.

\subsubsection{Pre-training Stage.}
The full CCMAN architecture, including the cross-attention modules, Transformer encoder, and projection head, is pre-trained on the 12 non-verbal-fluency cognitive assessment tasks described in Section~\ref{sec:data}. Training uses the same multimodal feature extraction pipeline and composite objective (Focal Loss, Supervised Contrastive Loss, and entropy regularisation; Section~\ref{subsec:trainingProc}), enabling the model to learn transferable cross-modal representations from diverse cognitive tasks before fine-tuning on verbal fluency classification.

\subsubsection{Fine-tuning Stage}
\label{subsec:finetuning}

Following pre-training, the learned weights are transferred to the verbal fluency tasks using a standard fine-tuning protocol. Task-agnostic representation-learning components are frozen to preserve the cross-modal representations learned during pre-training, while the task-specific projection and classification heads are updated.

\paragraph{Frozen Layers}
To prevent catastrophic forgetting, the following pre-trained components, explained in Table \ref{tab:finetune_config}, remain fixed during fine-tuning. 

\begin{itemize}

\item \textbf{Modality projection layers.}
Semantic ($\mathbf{s}_t$), acoustic ($\mathbf{a}_t$), and auxiliary ($\mathbf{o}_t$) features are projected into a shared 256-dimensional latent space,
\[
\tilde{\mathbf{s}}_t=W_s\mathbf{s}_t+\mathbf{b}_s,\qquad
\tilde{\mathbf{a}}_t=W_a\mathbf{a}_t+\mathbf{b}_a,\qquad
\tilde{\mathbf{o}}_t=W_o\mathbf{o}_t+\mathbf{b}_o.
\]

Freezing these layers preserves the learned multimodal feature alignment.

\item \textbf{Bidirectional cross-attention modules.}
Cross-modal interactions are computed as
\[
\mathbf{H}_{s\rightarrow a}=\mathrm{CrossAttn}(\tilde{\mathbf{s}},\tilde{\mathbf{a}}),\qquad
\mathbf{H}_{a\rightarrow s}=\mathrm{CrossAttn}(\tilde{\mathbf{a}},\tilde{\mathbf{s}}),
\]
capturing semantic--acoustic dependencies learned during pre-training.

\item \textbf{Transformer encoder.}
The fused sequence
\[
\mathbf{Z}=\mathrm{Transformer}(\mathbf{H})
\]
is also frozen, preserving the learned temporal representations.

\end{itemize}

\begin{table*}[!ht]
\centering
\scriptsize
\setlength{\tabcolsep}{5pt}
\renewcommand\arraystretch{0.7}
\caption{Classification performance (\textit{Macro} $F_1$, Precision, and Recall; mean $\pm$ standard deviation across cross-validation folds) for semantic and phonemic verbal fluency tasks. Results compare static and temporal baselines, the proposed CCMAN with incremental ablations, and external validation on the independent PROCESS-2 benchmark.}
\label{tab:fluency_results}
\begin{tabular}{l | l | l | l | l | c c c}
\toprule
\textbf{Task} & \textbf{Setup} & \textbf{Model} & \textbf{Best-performed Feature} & \textbf{Config} & \textbf{Macro $F_1$} & \textbf{Precision} & \textbf{Recall} \\
\midrule

\multicolumn{8}{c}{\textbf{Semantic Fluency Task}} \\
\midrule

\multirow{11}{*}{Binary}
& \multirow{4}{*}{Static}
& XGBoost & Acoustics (\egemaps{} \cite{eyben2015geneva} + \compare{} \cite{schuller2016interspeech, eyben2010opensmile}) & Standard & 0.66$\pm$0.04 & 0.66$\pm$0.03 & 0.68$\pm$0.03 \\
& & XGBoost & Sentence Embeddings (\SentenceBERT \cite{reimers2019sentence}, $D_s=384$) & Standard & 0.68$\pm$0.04 & 0.67$\pm$0.03 & 0.68$\pm$0.03 \\
& & XGBoost & Pause features ($\mathbf{p}_t \in \mathbb{R}^{4}$) & Standard & 0.65$\pm$0.03 & 0.66$\pm$0.03 & 0.66$\pm$0.03 \\
& & LLM & ASR-transcribed Text & Standard & \textit{0.75$\pm$0.03} & \textit{0.76$\pm$0.03} & \textit{0.76$\pm$0.03} \\

\cmidrule(lr){2-8}

& \multirow{7}{*}{Temporal}
& XGBoost & Sequential features ($X \in \mathbb{R}^{B \times T \times D}$) & Standard & 0.65$\pm$0.04 & 0.65$\pm$0.04 & 0.67$\pm$0.04 \\
\cmidrule(lr){3-8}
& & \multirow{5}{*}{CCMAN} & \multirow{5}{*}{Word tokens (\WV{}) + Embeddings (\SentenceBERT{})}
& Base & 0.75$\pm$0.03 & 0.75$\pm$0.03 & 0.76$\pm$0.03 \\
& & & & + Drift & 0.77$\pm$0.03 & 0.78$\pm$0.03 & 0.77$\pm$0.03 \\
& & & & + CrossAttn & 0.78$\pm$0.02 & 0.78$\pm$0.02 & 0.79$\pm$0.02 \\
& & & & + Entropy & 0.79$\pm$0.02 & 0.80$\pm$0.02 & 0.80$\pm$0.02 \\
& & & & + TF & \textbf{0.81$\pm$0.02} & \textbf{0.82$\pm$0.02} & \textbf{0.82$\pm$0.02} \\

\midrule

\multirow{11}{*}{Multiclass}
& \multirow{4}{*}{Static}
& XGBoost & Acoustics (\egemaps{} \cite{eyben2015geneva} + \compare{} \cite{schuller2016interspeech, eyben2010opensmile}) & Standard & 0.39$\pm$0.04 & 0.39$\pm$0.03 & 0.40$\pm$0.03 \\
& & XGBoost & Sentence Embeddings (\SentenceBERT \cite{reimers2019sentence}, $D_s=384$) & Standard & 0.39$\pm$0.04 & 0.40$\pm$0.03 & 0.40$\pm$0.03 \\
& & XGBoost & Pause features ($\mathbf{p}_t \in \mathbb{R}^{4}$) & Standard & 0.39$\pm$0.03 & 0.40$\pm$0.03 & 0.39$\pm$0.03 \\
& & LLM & ASR-transcribed Text & Standard & \textit{0.50$\pm$0.03} & \textit{0.51$\pm$0.03} & \textit{0.50$\pm$0.03} \\

\cmidrule(lr){2-8}

& \multirow{7}{*}{Temporal}
& XGBoost & Sequential features ($X \in \mathbb{R}^{B \times T \times D}$) & Standard & 0.41$\pm$0.04 & 0.42$\pm$0.04 & 0.42$\pm$0.04 \\
\cmidrule(lr){3-8}
& & \multirow{5}{*}{CCMAN} & \multirow{5}{*}{Word tokens (\Whisper{}) + Embeddings (\WordVec{})}
& Base & 0.50$\pm$0.03 & 0.51$\pm$0.03 & 0.51$\pm$0.03 \\
& & & & + Drift & 0.52$\pm$0.03 & 0.53$\pm$0.03 & 0.52$\pm$0.03 \\
& & & & + CrossAttn & 0.54$\pm$0.02 & 0.55$\pm$0.02 & 0.55$\pm$0.02 \\
& & & & + Entropy & 0.56$\pm$0.02 & 0.57$\pm$0.02 & 0.57$\pm$0.02 \\
& & & & + TF & \textbf{0.59$\pm$0.02} & \textbf{0.59$\pm$0.02} & \textbf{0.60$\pm$0.02} \\

\midrule
\multicolumn{8}{c}{\textbf{Phonemic Fluency Task}} \\
\midrule

\multirow{11}{*}{Binary}
& \multirow{4}{*}{Static}
& XGBoost & Acoustics (\egemaps{} \cite{eyben2015geneva} + \compare{} \cite{schuller2016interspeech, eyben2010opensmile}) & Standard & 0.65$\pm$0.04 & 0.66$\pm$0.03 & 0.67$\pm$0.03 \\
& & XGBoost & Sentence Embeddings (\SentenceBERT \cite{reimers2019sentence}, $D_s=384$) & Standard & 0.67$\pm$0.04 & 0.67$\pm$0.03 & 0.68$\pm$0.03 \\
& & XGBoost & Pause features ($\mathbf{p}_t \in \mathbb{R}^{4}$) & Standard & 0.65$\pm$0.03 & 0.66$\pm$0.03 & 0.65$\pm$0.03 \\
& & LLM & ASR-transcribed Text & Standard & \textit{0.72$\pm$0.03} & \textit{0.73$\pm$0.03} & \textit{0.72$\pm$0.03} \\

\cmidrule(lr){2-8}

& \multirow{7}{*}{Temporal}
& XGBoost & Sequential features ($X \in \mathbb{R}^{B \times T \times D}$) & Standard & 0.64$\pm$0.04 & 0.65$\pm$0.04 & 0.67$\pm$0.04 \\
\cmidrule(lr){3-8}
& & \multirow{5}{*}{CCMAN} & \multirow{5}{*}{Word tokens (\WV{}) + Embeddings (\FastText{})}
& Base & 0.70$\pm$0.03 & 0.70$\pm$0.03 & 0.71$\pm$0.03 \\
& & & & + Drift & 0.71$\pm$0.03 & 0.70$\pm$0.03 & 0.72$\pm$0.03 \\
& & & & + CrossAttn & 0.74$\pm$0.02 & 0.75$\pm$0.02 & 0.75$\pm$0.02 \\
& & & & + Entropy & 0.75$\pm$0.02 & 0.75$\pm$0.02 & 0.76$\pm$0.02 \\
& & & & + TF & \textbf{0.77$\pm$0.02} & \textbf{0.77$\pm$0.02} & \textbf{0.77$\pm$0.02} \\

\midrule

\multirow{11}{*}{Multiclass}
& \multirow{4}{*}{Static}
& XGBoost & Acoustics (\egemaps{} \cite{eyben2015geneva} + \compare{} \cite{schuller2016interspeech, eyben2010opensmile}) & Standard & 0.37$\pm$0.04 & 0.38$\pm$0.03 & 0.37$\pm$0.03 \\
& & XGBoost & Sentence Embeddings (\SentenceBERT \cite{reimers2019sentence}, $D_s=384$) & Standard & 0.39$\pm$0.04 & 0.40$\pm$0.03 & 0.39$\pm$0.03 \\
& & XGBoost & Pause features ($\mathbf{p}_t \in \mathbb{R}^{4}$) & Standard & 0.36$\pm$0.03 & 0.36$\pm$0.03 & 0.37$\pm$0.03 \\
& & LLM & ASR-transcribed Text & Standard & \textit{0.46$\pm$0.03} & \textit{0.46$\pm$0.03} & \textit{0.47$\pm$0.03} \\

\cmidrule(lr){2-8}

& \multirow{7}{*}{Temporal}
& XGBoost & Sequential features ($X \in \mathbb{R}^{B \times T \times D}$) & Standard & 0.40$\pm$0.04 & 0.39$\pm$0.04 & 0.42$\pm$0.04 \\
\cmidrule(lr){3-8}
& & \multirow{5}{*}{CCMAN} & \multirow{5}{*}{Word tokens (\Whisper{}) + Embeddings (\FastText{})}
& Base & 0.47$\pm$0.03 & 0.47$\pm$0.03 & 0.47$\pm$0.03 \\
& & & & + Drift & 0.49$\pm$0.03 & 0.49$\pm$0.03 & 0.50$\pm$0.02 \\
& & & & + CrossAttn & 0.50$\pm$0.02 & 0.50$\pm$0.02 & 0.50$\pm$0.02 \\
& & & & + Entropy & 0.51$\pm$0.02 & 0.50$\pm$0.01 & 0.52$\pm$0.01 \\
& & & & + TF & \textbf{0.53$\pm$0.02} & \textbf{0.52$\pm$0.01} & \textbf{0.54$\pm$0.01} \\


\midrule
\midrule
\multicolumn{7}{c}{\textbf{Public benchmark dataset (PROCESS-2) - SFT}} \\
\midrule
Binary & \multirow{2}{*}{Temporal} & \multirow{2}{*}{CCMAN} & Word tokens (\Whisper{}) + Embeddings (\WordVec{}) & \multirow{2}{*}{Base + TF} & 0.75 (+4\%) & 0.76 & 0.75 \\
Multiclass &  &  & Word tokens (\WV{}) + Embeddings (\FastText{}) &  & \textbf{0.56 (+9\%)} & 0.56 & 0.57 \\

\midrule
\multicolumn{7}{c}{\textbf{Public benchmark dataset (PROCESS-2) - PFT}} \\
\midrule
Binary & \multirow{2}{*}{Temporal} & \multirow{2}{*}{CCMAN} & Word tokens (\Whisper{}) + Embeddings (\WordVec{}) & \multirow{2}{*}{Base + TF} & 0.76 (+1\%) & 0.76 & 0.77 \\
Multiclass &  &  & Word tokens (\Whisper{}) + Embeddings (\FastText{}) &  & 0.59 (+2\%) & 0.59 & 0.59 \\

\bottomrule
\end{tabular}
\end{table*}

\paragraph{Trainable Layers}
The task-specific components are fine-tuned as follows.

\begin{itemize}

\item \textbf{Projection head.}
The pooled representation is projected into a 64-dimensional embedding,
\[
\mathbf{h}=f_p(\mathbf{z})=\mathrm{MLP}_p(\mathbf{z}),
\]
which is $\ell_2$-normalised and optimised using supervised contrastive learning.

\item \textbf{Classification head.}
Predictions are obtained from
\[
\hat{\mathbf{y}}=f_c(\mathbf{z})=\mathrm{MLP}_c(\mathbf{z}),
\]
where the final linear layer is re-initialised for the target task ($C=2$ or $3$), while the remaining classifier layers are fine-tuned from their pre-trained weights.

\end{itemize}
\paragraph{Training Strategy.}
Both the projection head and the classification head are unfrozen and updated together with a lower learning rate ($1\times10^{-5}$) compared to the pre-training phase ($1\times10^{-4}$). This freeze–adapt strategy serves three purposes: (i) it prevents catastrophic forgetting of the generalisable representations learned during pre-training; (ii) it reduces the risk of overfitting to the smaller fine-tuning dataset; and (iii) it significantly lowers computational cost, as only a small subset of parameters (approximately 15\% of the total) are updated.

Table~\ref{tab:finetune_config} summarises the fine-tuning configuration. The same 5-fold subject-level cross-validation protocol is used, with results reported in Table~\ref{tab:fluency_results}. We denote \textit{Standard Fine-Tuning} (SFT) as fine-tuning without pre-training (training from scratch), and \textit{Pre-trained Fine-Tuning} (PFT) as the full transfer learning pipeline described above.

\subsection{Training Procedure}
\label{subsec:trainingProc}

We use 5-fold subject-level cross-validation and train end-to-end with AdamW. Class imbalance is addressed using class-weighted Focal Loss ($\gamma=2$) \cite{pahar2024nanocnn, lin2017focal, paharliudatapaper}. 
Supervised Contrastive Loss (temperature $=0.07$) is applied to projection outputs (weight $=0.3$) \cite{khosla2020supervised}. 
Cross-modal attention entropy regularisation (weight $=0.01$) promotes informative alignments \cite{mohamed2015variational}. 
These hyperparameters were optimised during the model-training protocol, and performance is reported using macro $F_1$-score, macro precision, and macro recall in Section \ref{sec:results}.


\section{Results}
\label{sec:results}

Table~\ref{tab:fluency_results} reports classification performance for static (aggregated and adapted in \cite{pahar2025cognospeak, pahar2025mutlimodalfusion}) and temporal (sequential and presented in this study) models on semantic and phonemic fluency tasks under binary and multiclass settings. Static baselines achieved moderate binary performance ($F_1$: 0.65--0.75), with LLM-based (\bart{} \cite{lewis2019bart}, \roberta{} \cite{liu2019roberta} and \distilbert{} \cite{sanh2019distilbert}) text features performing best (Semantic: 0.75; Phonemic: 0.72). Performance dropped in multiclass classification ($F_1$: 0.36--0.50), showing limited sensitivity to finer-grained distinctions; adding handcrafted temporal descriptors gave only marginal gains.

The proposed CCMAN consistently outperformed all baselines. Semantic fluency $F_1$ reached 0.81 (binary) and 0.59 (multiclass), representing improvements of 6\% and 9\% over the strongest static baselines, respectively. Phonemic fluency achieved corresponding scores of 0.77 (binary) and 0.53 (multiclass), improving upon the best static models by 5\% and 7\%. Improvements were largest in multiclass classification, highlighting the benefit of cross-modal temporal modelling for capturing subtle differences between HC, MCI and dementia. Furthermore, CCMAN consistently exhibited lower standard deviations across cross-validation folds, indicating improved robustness and generalisation. Ablation analysis confirmed the contribution of each component: the base architecture already surpassed the static baselines, semantic drift modelling yielded consistent gains, bidirectional cross-attention produced the largest performance improvement, and entropy-based temporal modelling further enhanced robustness, resulting in the best overall performance.

To evaluate external generalisability, the best-performing CCMAN was further evaluated on the independent PROCESS-2 benchmark without architectural modifications. 
We note that although PROCESS-2 was collected using the \cognospeak{} platform, there was no overlap between PROCESS-2 and the data explained in Table \ref{tab:dataset_summary_asr_group}. 
CCMAN achieved Macro-$F_1$ scores of 0.75 (binary) and 0.56 (multiclass) on the semantic fluency task, improving upon the published baselines by 4\% and 9\%, respectively. On the phonemic fluency task, it achieved Macro-$F_1$ scores of 0.76 (binary) and 0.59 (multiclass), corresponding to improvements of 1\% and 2\%. The consistent performance gains across both datasets demonstrate that the proposed temporal cross-modal representations learn transferable cognitive speech biomarkers that generalise beyond the CognoSpeak cohort.


\section{Discussion}

\subsection{Global Instability}


The best-performing semantic fluency model under multiclass classification (Table~\ref{tab:fluency_results}) was selected for biomarker interpretation using the proposed global and progressive temporal instability measures, with statistics averaged across the test folds. Table~\ref{tab:feature_stats} summarises the statistical comparison of the global instability biomarkers across HC, individuals with MCI, and dementia. Kruskal--Wallis tests identified significant group differences for both \textit{Semantic Drift Variance} ($H=18.65$, FDR-adjusted $p=2.68\times10^{-4}$) and \textit{Pause Variance} ($H=17.00$, FDR-adjusted $p=3.05\times10^{-4}$), whereas \textit{Drift Mean} was not significantly different across groups ($p=0.12$). Median values increased monotonically across groups for semantic drift variance (HC = 0.0323, MCI = 0.0368, dementia = 0.0477) and pause variance (HC = 0.0045, MCI = 0.0058, dementia = 0.0060). Post-hoc Mann--Whitney U tests with Bonferroni correction confirmed that both features were significantly higher in MCI and dementia than in HC (drift variance: $p_{\mathrm{bonf}} = 1.01\times10^{-2}$ and $1.36\times10^{-3}$; pause variance: $p_{\mathrm{bonf}} = 2.46\times10^{-2}$ and $1.41\times10^{-3}$), while MCI--dementia differences were not statistically significant ($p_{\mathrm{bonf}} = 0.35$ and $0.43$, respectively). Drift mean also increased numerically from HC (0.1575) to MCI (0.1707) to dementia (0.1755), but this trend did not reach significance.

\begin{table*}[h]
\centering
\scriptsize
\setlength{\tabcolsep}{2pt}
\renewcommand\arraystretch{0.7}
\caption{Group differences in global instability features across three classes. 
Global differences were assessed via Kruskal--Wallis H tests with Benjamini--Hochberg (FDR) correction. 
Significant results were further analysed using post-hoc Mann--Whitney U tests with Bonferroni correction. 
Group labels: 0 = HC, 1 = MCI, 2 = Dementia.}
\label{tab:feature_stats}
\begin{tabular}{@{}l c c c c c@{}}
\toprule
\textbf{Feature} & \textbf{Median (HC/MCI/Dem)} & \textbf{Kruskal H} & \textbf{FDR-adj. $p$} & \textbf{Pairwise} & \textbf{Post-hoc $p_{bonf}$} \\ \midrule
Drift Mean       & 0.1575 / 0.1707 / 0.1755 & 4.24  & $1.20 \times 10^{-1}$ & ---    & --- \\ \midrule
Drift Variance   & 0.0323 / 0.0368 / 0.0477 & 18.65 & \textbf{2.68 $\times 10^{-4}$***} & 0 vs 1 & $1.01 \times 10^{-2}$* (1>0) \\
                 &                          &       &                       & 0 vs 2 & $1.36 \times 10^{-3}$** (2>0) \\
                 &                          &       &                       & 1 vs 2 & $3.46 \times 10^{-1}$ (ns) \\ \midrule
Pause Variance   & 0.0045 / 0.0058 / 0.0060 & 17.00 & \textbf{3.05 $\times 10^{-4}$***} & 0 vs 1 & $2.46 \times 10^{-2}$* (1>0) \\
                 &                          &       &                       & 0 vs 2 & $1.41 \times 10^{-3}$** (2>0) \\
                 &                          &       &                       & 1 vs 2 & $4.25 \times 10^{-1}$ (ns) \\ \bottomrule
\end{tabular}
\begin{flushleft}
\footnotesize{Note: Significance levels: *** $p < 0.001$, ** $p < 0.01$, * $p < 0.05$, (ns) non-significant. FDR: False Discovery Rate; $p_{bonf}$: Bonferroni-corrected p-value.}
\end{flushleft}
\end{table*}

\begin{figure}[ht]
\centering
\includegraphics[width=\linewidth]{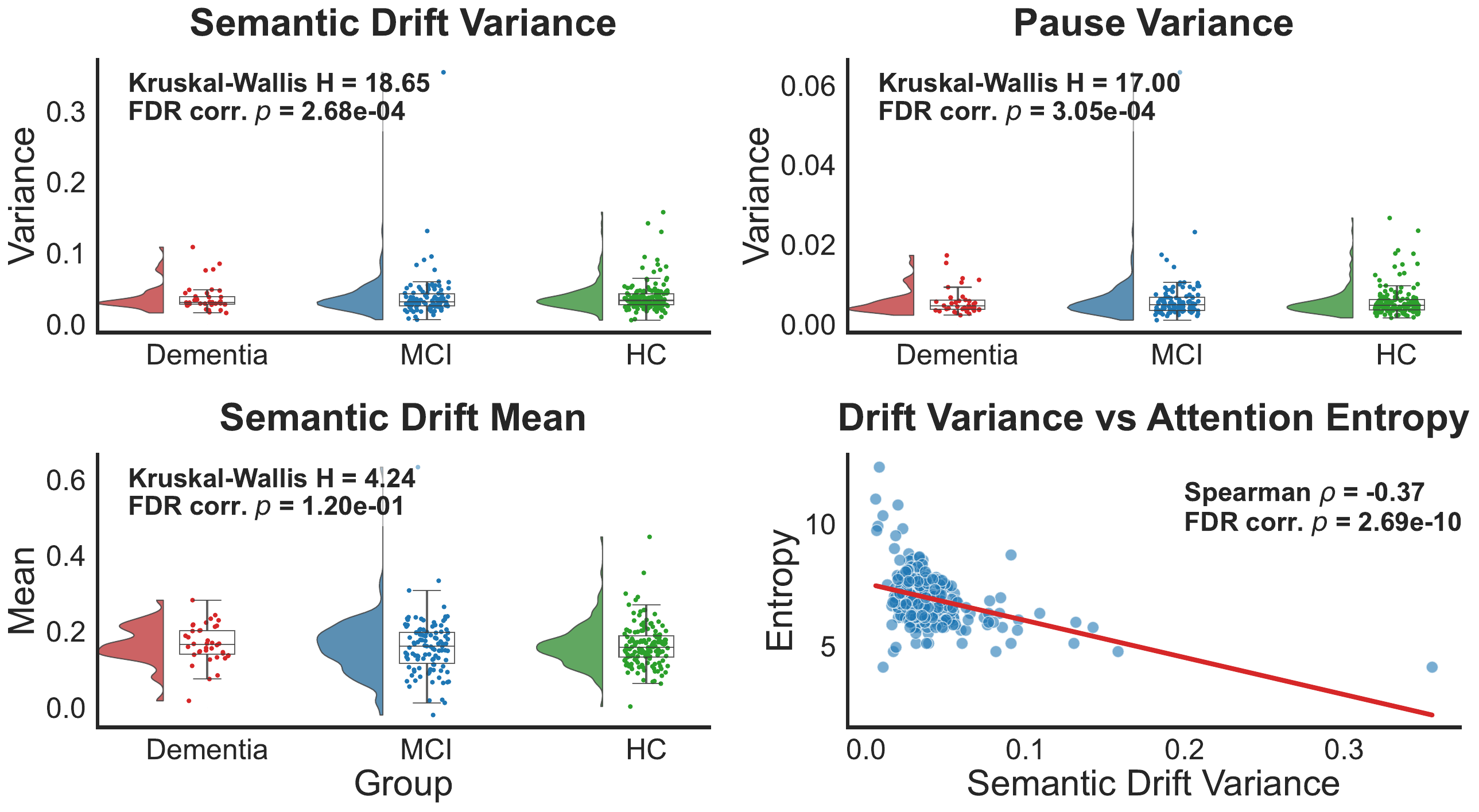}
\caption{
\textbf{Group differences in global instability metrics}. 
Top-left: Drift variance shows significant increases across classes. 
Top-right: Pause variance also increases significantly. 
Bottom-left: Drift mean does not differ significantly, indicating that instability rather than absolute semantic shift differentiates groups. 
Bottom-right: A significant negative correlation between drift variance and attention entropy suggests that increased semantic instability is associated with more concentrated model attention.
}
\label{fig:instability}
\end{figure}

Figure~\ref{fig:instability} visualises these group differences. The absence of significant changes in \textit{Drift Mean} suggests that cognitive impairment is characterised by increased variability in semantic progression rather than larger absolute semantic shifts. Clinically, consistent with the observed median increases, higher \textit{Semantic Drift Variance} indicates less structured transitions between semantic subcategories during verbal fluency, while increased \textit{Pause Variance} reflects irregular lexical retrieval timing, with fluent word production interspersed by prolonged hesitations rather than uniform slowing. Together, these findings suggest greater temporal instability in semantic search and executive control among cognitively impaired individuals, consistent with previous evidence of disrupted semantic organisation and executive dysfunction in neurodegenerative disease \cite{troyer1997clustering,grossman2003neural}. Although dementia showed the highest median values, the lack of a significant MCI--dementia contrast suggests that these instability markers may be more sensitive to the presence of cognitive impairment than to its severity. Furthermore, the negative association between semantic drift variance and attention entropy indicates that as semantic retrieval becomes increasingly unstable, CCMAN concentrates its cross-modal attention on fewer informative speech segments, suggesting a compensatory mechanism for handling cognitive instability.

\subsection{Temporal Dynamics}

\begin{figure}[ht]
\centering
\includegraphics[width=\linewidth]{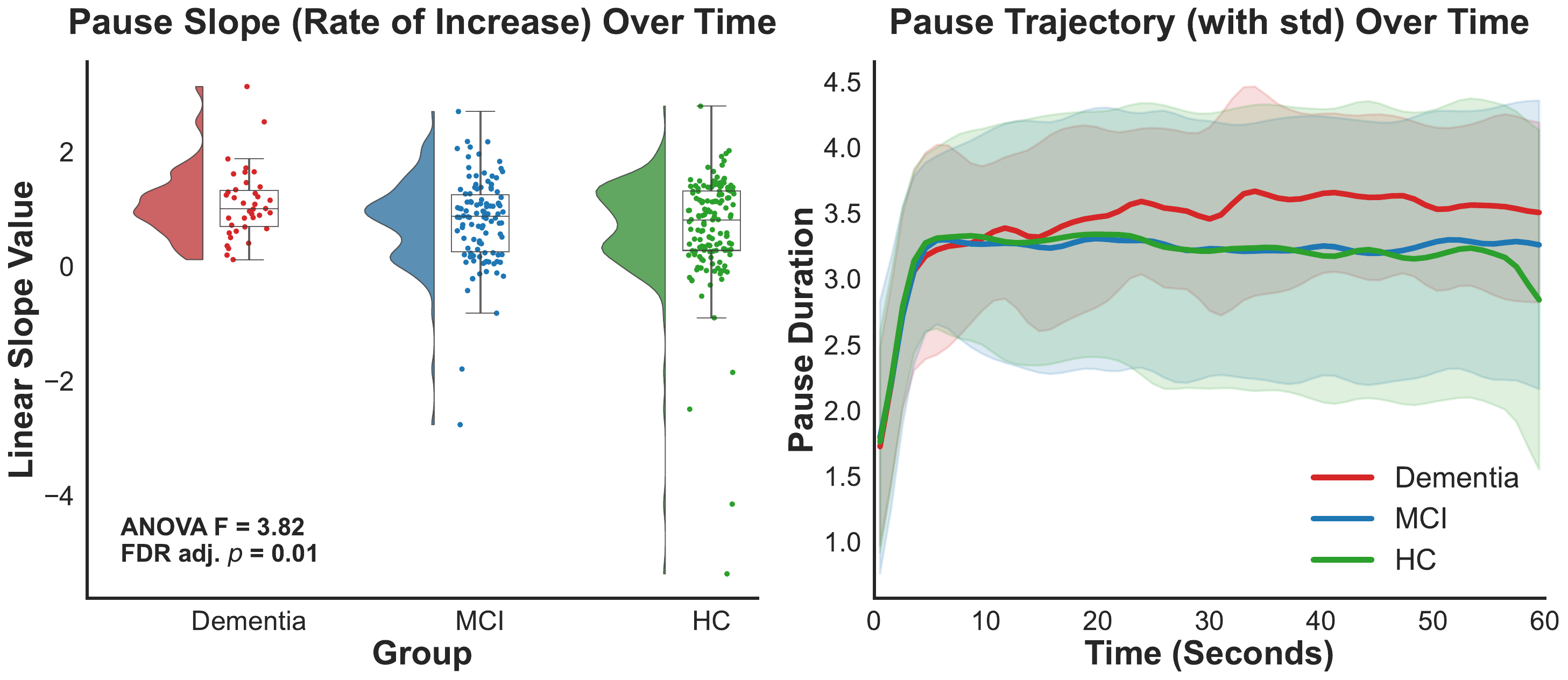}
\caption{\textbf{Temporal dynamics between diagnoses.}
Left: Dementia exhibits significantly steeper positive slopes relative to MCI and HC, indicating a greater progressive increase in pause duration over time.
Right: Pause duration increases progressively, with the steepest slope observed in dementia.}
\label{fig:temporal}
\end{figure}

Temporal slope analysis (Table~\ref{tab:slope_stats_final}, Figure~\ref{fig:temporal}) revealed a significant group effect only for \textit{Pause} ($F=3.82$, FDR-adjusted $p=0.01$). Pause duration increased progressively from HC to MCI to Dementia, with the steepest slope in Dementia, indicating accumulating retrieval effort over discourse. Clinically, this pattern suggests progressive depletion of lexical search resources rather than a static retrieval deficit, consistent with reduced cognitive endurance in dementia. The absence of slope effects for \textit{Semantic Drift}, \textit{Gate}, and \textit{Entropy} indicates that temporal escalation is specific to speech timing rather than broader cross-modal instability dynamics.

\begin{table}[ht]
\centering
\scriptsize
\setlength{\tabcolsep}{4pt}
\renewcommand\arraystretch{1.2}
\caption{One-way ANOVA slope analysis of temporal features across diagnostic groups (HC, MCI, Dementia). 
}
\label{tab:slope_stats_final}
\begin{tabular}{@{}lccc@{}}
\toprule
\textbf{Feature} & \textbf{F-value} & \textbf{FDR-adj. $p$} & \textbf{Mean Slope (HC $\rightarrow$ MCI $\rightarrow$ Dem)} \\ \midrule
Pause            & 3.82             & 0.01*                 & 0.68 $\rightarrow$ 0.80 $\rightarrow$ 1.09            \\
Gate             & 0.92             & 0.76                 & 0.00 $\rightarrow$ 0.00 $\rightarrow$ 0.00            \\
Entropy          & 0.39             & 0.76                 & 0.00 $\rightarrow$ 0.00 $\rightarrow$ 0.00            \\
Semantic Drift   & 0.27             & 0.76                 & -0.22 $\rightarrow$ -0.23 $\rightarrow$ -0.25         \\ \bottomrule
\end{tabular}
\begin{flushleft}
\footnotesize{Note: $p$-values are adjusted using the Benjamini-Hochberg False Discovery Rate (FDR) procedure. Mean slope values illustrate the progression from HC to MCI to Dementia.}
\end{flushleft}
\end{table}

\subsection{Integrating Global and Temporal Instability}

Integrating global and temporal instability reveals complementary biomarkers of cognitive decline. Global variance measures capture overall semantic instability, whereas pause slopes reflect progressively increasing retrieval effort over time. Dementia exhibits both elevated global instability and temporal escalation, while MCI shows increased variability without a pronounced temporal trend. Together, these findings support a \textit{dynamic speech biomarker} framework that characterises both global fluctuations and evolving speech effort. 

\subsection{Limitations}
As the tasks in \cognospeak{} differ from those in corpora such as \dementiabank{} \cite{becker1994natural}, cross-dataset evaluation was limited to the PROCESS-2 benchmark \cite{tao2025PROCESS, pahar2026PROCESS2}. 
Despite a relatively high ASR error rate, CCMAN consistently outperformed both static and temporal baselines and generalised to the independent PROCESS-2 benchmark, suggesting that the proposed multimodal transfer learning framework is robust to transcription errors encountered in real-world clinical speech.
To facilitate reproducibility, a subset of the dataset together with the complete CCMAN framework, including feature extraction, model architecture, and temporal analysis, has been released publicly \cite{Pahar2026CCMAN}.


\section{Conclusion}

We presented CCMAN, a Cognitive Instability-Aware Cross-Modal Attention Network, for detecting cognitive decline from semantic and phonemic verbal fluency speech. Unlike conventional approaches that rely on recording-level aggregated features, CCMAN models word-level temporal dynamics by integrating semantic, acoustic, linguistic, pause, and semantic drift information through cross-modal attention and transformer-based sequence modelling. By pretraining on multiple cognitive assessment tasks and subsequently fine-tuning on verbal fluency, the proposed transfer learning framework learns transferable multimodal representations that consistently outperform strong static and temporal baselines in both binary and multiclass classification.
Beyond improved predictive performance, CCMAN provides interpretable insights into the cognitive processes underlying speech production. 
Statistical analyses showed that semantic drift variance and pause variance, but not mean semantic drift, were significantly elevated in both MCI and dementia relative to healthy controls, while pause duration increased progressively over the task with the steepest slope in dementia, revealing increasing lexical retrieval effort throughout the verbal fluency task.
Furthermore, evaluation on the independent PROCESS-2 benchmark demonstrated that the learned representations generalise across datasets collected under different protocols, highlighting the robustness of the proposed transfer learning strategy.

Overall, this work establishes temporal speech instability as a measurable and interpretable biomarker and demonstrates that our cross-modal sequence modelling provides a principled and computationally efficient framework for scalable, speech-based early cognitive impairment detection.
Future work will investigate longitudinal modelling across repeated assessments and domain adaptation to enhance generalisability across populations and recording conditions.

\section{Acknowledgments}

This research was partly funded by the NIHR Sheffield Biomedical Research Centre (BRC), and the NIHR202911 award under the NIHR i4i programme. The views expressed are those of the authors and not necessarily those of the NHS, the NIHR or the Department of Health and Social Care (DHSC).
Ethical approval for the collection of the data analysed in this study was granted by the NRES Committee South West-Central Bristol (REC number 16/LO/0737).

For the purpose of open access, the author has applied a Creative Commons Attribution (CC BY) licence to any Author Accepted Manuscript version arising. 
All participants took part in this study voluntarily and provided their consent to use their data for research purposes.

\bibliographystyle{IEEEtran}
\bibliography{mybib}

\end{document}